\documentclass[a4paper,12pt]{article}
\usepackage{amsfonts}

\usepackage{graphicx}

\input{tcilatex}

\begin{document}

%%      figures
%
% Fig.1 YangB.eps
% Fig.2 boot.eps
% Fig.3 bootB.eps
%
%%%%%%%%%%%%%

\setcounter{page}{0} \topmargin0pt \oddsidemargin5mm \renewcommand{%
\thefootnote}{\fnsymbol{footnote}} \newpage \setcounter{page}{0} 
\begin{titlepage}
\begin{flushright}
Berlin Sfb288 Preprint  \\
hep-th/0201142\\
2-nd revised version
\end{flushright}
\vspace{0.2cm}
\begin{center}
{\large {\bf On the absence of simultaneous reflection and 
transmission in integrable impurity systems} }

\vspace{0.8cm}
{\large  O.A.~Castro-Alvaredo$^*$,  A.~Fring$^*$ and  F.~G\"ohmann$^\dag$ }

\vspace{0.2cm}
{$^*$Institut f\"ur Theoretische Physik, 
Freie Universit\"at Berlin,\\
Arnimallee 14, D-14195 Berlin, Germany\\
$^\dag$Theoretische Physik I, Universit\"at Bayreuth, D-95440 Bayreuth, Germany }
\end{center}
\vspace{0.5cm}
 
\renewcommand{\thefootnote}{\arabic{footnote}}
\setcounter{footnote}{0}

\begin{abstract}
We establish that the Yang-Baxter equations in the presence of an impurity 
can in general only admit solutions of simultaneous transmission and 
reflection when the transmission and reflection amplitudes commute in the 
defect degrees of freedom with an additional exchange of the corresponding 
rapidities. In the absence of defect degrees of freedom we show in complete 
generality, that the only exceptions to this are theories which possess 
rapidity independent bulk scattering matrices. In particular bulk theories
with diagonal scattering matrices, can only be 
the free Boson and Fermion, the Federbush model and their generalizations. 
These anyonic solutions do not admit the possibility 
of excited impurity states.
\end{abstract}
\vfill{ \hspace*{-9mm}
\begin{tabular}{l}
\rule{6 cm}{0.05 mm}\\
Olalla@physik.fu-berlin.de\\
Fring@physik.fu-berlin.de\\
Frank.Goehmann@uni-bayreuth.de
\end{tabular}}
\end{titlepage}
\newpage 

\section{Introduction}

Integrable quantum field theories in 1+1 space-time dimensions in the
presence of a boundary have received a considerable amount of attention in
recent years. One of the central aims is to find explicit solutions to the
consistency equations in the presence of a boundary, which result as a
consequence of factorizability, namely the Yang-Baxter equation \cite
{Chered,Skly}, the bootstrap equation \cite{FK} and also crossing \cite
{GZ,Hou}. Explicit solutions are known for various theories, such as affine
Toda field theories with real \cite{FK2,Sasaki,CARS,Riva} and purely
imaginary coupling \cite{Georg,GG}, (in particular the sine-Gordon model 
\cite{GZ,Ghosh1,AW,BPT,BPTT} and its supersymmetric version \cite{IOZ,Nepo}%
), the Gross-Neveu model \cite{deMM}, $N=1$ \cite{MS} and $N=2$ \cite{Nepo2}
supersymmetric theories, the nonlinear sigma models \cite{Ghosh,MM,MacK} and
theories with infinite resonance states \cite{MBR}.

Part of the motivation for this great interest is resultig from the fact
that boundaries play a natural role in string theory. In the context of
condensed matter physics, boundaries allow for instance the description of
non-trivial constrictions in quantum wires. In order to understand realistic
materials, it is in addition further important to investigate the effects of
impurities (defects, inhomogeneities). For this latter situation much less
is known at present. Besides the purely reflecting case, which is equivalent
to the aforementioned boundary problem, there exist some solutions for
purely transmitting impurities \cite{Konik}. However, hitherto only few
examples are known for the situation of simultaneously occurring reflection
and transmission \cite{DMS,Konik2,CF10}. All these examples studied so far
are related either to the free Fermion or Boson.

In \cite{DMS} an argument was provided, which manifests that integrable
parity invariant impurity systems with a diagonal bulk S-matrix, apart from $%
S=\pm 1$, do not allow simultaneously non-trivial reflection and
transmission amplitudes. In this note we address the question, whether the
set of possible bulk theories, which admit such a behaviour of the impurity,
can be enlarged when the corresponding S-matrix is taken to be non-diagonal
and parity is allowed to be broken. Without making any assumptions it will
turn out that non-diagonal bulk scattering theories do not admit the
possibility of simultaneous reflection and transmission on the defect when
integrability of the theory is maintained and degrees of freedom in the
defect are absent. When allowing additional degrees of freedom in the
impurity one can only have simultaneous transmission and reflection for
non-diagonal bulk theories when the assumption (\ref{ass}) does not hold. We
show that when allowing parity breaking the set of possible bulk theories with
such a behaviour of the defect can be slighly enlarged to those which are of
anyonic type. We demonstrate that, whenever reflection and transmission
occur at the same time, the defect can only have one degree of freedom.

\section{Defect Yang-Baxter equations}

Integrability is, as usual in this context, identified with the
factorization of the n-particle scattering matrix into two-particles ones.
Many of the properties, these two-particle scattering matrices have to
satisfy, result from the exploitation of the associativity of the so-called
Faddeev-Zamolodchikov (FZ) algebra \cite{FZ}. This also holds in the
presence of a boundary \cite{Chered,Skly,FK} or an impurity \cite{DMS},
which formally can be associated to an element of the algebra with zero
rapidity. We briefly want to recall this derivation, by following largely 
\cite{DMS}, with the difference that we also allow additional degrees of
freedom in the inhomogeneity, corresponding to possible excited impurity
states, and pay attention to parity. The latter means in particular, that
amplitudes may be different when particles hit the defect from the left or
from the right, a property known for instance in the context of lattice
integrable models, see e.g. \cite{Kadar}. Indicating particle types by
Latin and degrees of freedom of the impurity by Greek letters, the
``braiding'' relations of creation operators $Z_{i}(\theta )$ of a particle
of type $i$ with rapidity $\theta $ and defect operators $Z_{\alpha }$ in
the state $\alpha $ can be written as 
\begin{eqnarray}
Z_{i}(\theta _{1})Z_{j}(\theta _{2}) &=&S_{ij}^{kl}(\theta _{1}-\theta
_{2})Z_{k}(\theta _{2})Z_{l}(\theta _{1}),  \label{Z1} \\
Z_{i}(\theta )Z_{\alpha } &=&R_{i\alpha }^{j\beta }(\theta )Z_{j}(-\theta
)Z_{\beta }+T_{i\alpha }^{j\beta }(\theta )Z_{\beta }Z_{j}(\theta )\,, \\
Z_{\alpha }Z_{i}(\theta ) &=&\tilde{R}_{i\alpha }^{j\beta }(-\theta
)Z_{\beta }Z_{j}(-\theta )+\tilde{T}_{i\alpha }^{j\beta }(-\theta
)Z_{j}(\theta )Z_{\beta }.  \label{Z2}
\end{eqnarray}
We employed Einstein's sum convention, that is we assume sums over doubly
occurring indices. The left/right reflection and transmission amplitudes are
denoted by $R/\tilde{R}$ and $T/\tilde{T}$, respectively. We suppress the
explicit mentioning of the dependence of $Z_{\alpha }$ on the position in
space and assume that it is included in $\alpha $. For the treatment of a
single defect this is not relevant anyhow, but it becomes of course
important when considering multiple defects.

The algebra (\ref{Z1})-(\ref{Z2}) can be used to derive various relations
amongst the scattering amplitudes. Using them twice leads to the constraints 
\begin{eqnarray}
S_{ij}^{kl}(\theta )S_{kl}^{mn}(-\theta ) &=&\delta _{i}^{m}\delta _{j}^{n},
\label{U1} \\
R_{i\alpha }^{j\beta }(\theta )R_{j\beta }^{k\gamma }(-\theta )+T_{i\alpha
}^{j\beta }(\theta )\tilde{T}_{j\beta }^{k\gamma }(-\theta ) &=&\delta
_{i}^{k}\delta _{\alpha }^{\gamma },  \label{U2} \\
R_{i\alpha }^{j\beta }(\theta )T_{j\beta }^{k\gamma }(-\theta )+T_{i\alpha
}^{j\beta }(\theta )\tilde{R}_{j\beta }^{k\gamma }(-\theta ) &=&0\,.
\label{U3}
\end{eqnarray}
The same equations also hold after performing a parity transformation, that
is for $R\leftrightarrow \tilde{R}$ and $T\leftrightarrow \tilde{T}$ in (\ref
{U2})-(\ref{U3}).

The Yang-Baxter equations are derived as usual by exploiting the
associativity of the ZF-algebra. Commencing with an initial state of the
form $Z_{i}(\theta _{1})Z_{j}(\theta _{2})Z_{\alpha }$ and commuting in the
order as depicted in figure 1 (the picture is to be read as equality, in the
sense that the two scattering events are equal and the part in the middle of
the defects serves as the income on the right and as the outcome for the
left defect scattering process), we obtain the defect Yang-Baxter equations
by reading off the coefficients from the linear independent asymptotic
states of the form $Z_{i}(-\theta _{1})Z_{j}(-\theta _{2})Z_{\alpha }$, $%
Z_{i}(-\theta _{1})Z_{\alpha }Z_{j}(\theta _{2})$, $Z_{i}(-\theta
_{2})Z_{\alpha }Z_{j}(\theta _{1})$ and $Z_{\alpha }Z_{j}(\theta
_{2})Z_{i}(\theta _{1})$ as 
\begin{eqnarray}
S_{ij}^{kl}(\theta _{12})R_{l\alpha }^{m\beta }(\theta _{1})S_{km}^{np}(\hat{%
\theta}_{12})R_{p\beta }^{t\gamma }(\theta _{2}) &=&R_{j\alpha }^{l\beta
}(\theta _{2})S_{il}^{mp}(\hat{\theta}_{12})R_{p\beta }^{k\gamma }(\theta
_{1})S_{mk}^{nt}(\theta _{12}),  \label{YB1} \\
S_{ij}^{kl}(\theta _{12})R_{l\alpha }^{m\beta }(\theta _{1})S_{km}^{np}(\hat{%
\theta}_{12})T_{p\beta }^{t\gamma }(\theta _{2}) &=&T_{j\alpha }^{t\beta
}(\theta _{2})R_{i\beta }^{n\gamma }(\theta _{1}),  \label{YB2} \\
S_{ij}^{kl}(\theta _{12})T_{l\alpha }^{m\beta }(\theta _{1})R_{k\beta
}^{p\gamma }(\theta _{2}) &=&R_{j\alpha }^{l\beta }(\theta _{2})S_{il}^{pn}(%
\hat{\theta}_{12})T_{n\beta }^{m\gamma }(\theta _{1}),  \label{YB3} \\
S_{ij}^{kl}(\theta _{12})T_{l\alpha }^{m\beta }(\theta _{1})T_{k\beta
}^{p\gamma }(\theta _{2}) &=&T_{j\alpha }^{l\beta }(\theta _{2})T_{i\beta
}^{k\gamma }(\theta _{1})S_{kl}^{pm}(\theta _{12}).  \label{YB4}
\end{eqnarray}
We abbreviated here the rapidity sum $\hat{\theta}_{12}=\theta _{1}+\theta
_{2}$ and difference $\theta _{12}=\theta _{1}-\theta _{2}$. In the absence
of degrees of freedom in the impurity, the first relation (\ref{YB1}) was
originally obtained in \cite{Chered,Skly}, whereas (\ref{YB2})-(\ref{YB4}),
apart from a few obvious typos in the indices, were first derived in \cite
{DMS}. A systematic investigation with the addition of degrees of freedom in
the boundary for the purely reflecting case was initiated in \cite{KS}. Note
that when multiplying the equation (\ref{YB3}) by $S(\theta _{21})$ from the
left, it becomes identical to equation (\ref{YB2}) upon using the unitarity
relation (\ref{U1}) and a subsequent exchange of $\theta _{1}$ and $\theta
_{2}$. Thus, we only need to treat three independent equations.

\begin{center}
\includegraphics[width=11.2cm,height=7.77cm]{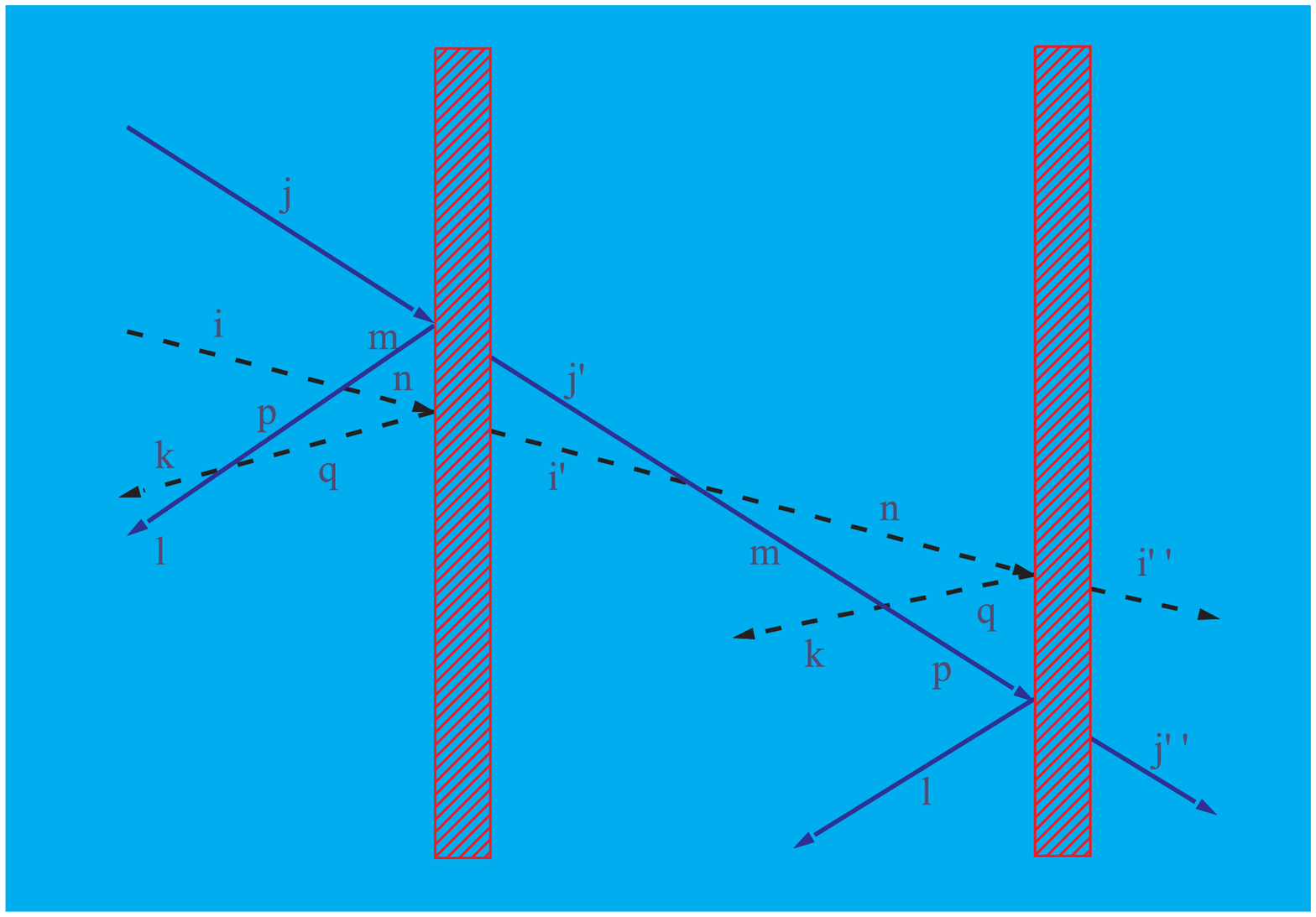}
\end{center}

\noindent {\small Figure 1: Defect Yang-Baxter equations. }

Starting with an initial state in a different order leads to non-equivalent
sets of equations. For instance taking $Z_{\alpha }Z_{i}(\theta
_{1})Z_{j}(\theta _{2})$ as the initial state simply leads to the same
equations as (\ref{YB1})-(\ref{YB4}) with $R\leftrightarrow \tilde{R}$ and $%
T\leftrightarrow \tilde{T}$. Commencing on the other hand with $Z_{i}(\theta
_{1})Z_{\alpha }Z_{j}(\theta _{2})$ and reading off the coefficients from
the linear independent asymptotic states of the form $Z_{i}(-\theta
_{1})Z_{\alpha }Z_{j}(-\theta _{2})$, $Z_{i}(-\theta _{1})Z_{j}(\theta
_{2})Z_{\alpha }$, $Z_{j}(\theta _{2})Z_{\alpha }Z_{i}(\theta _{1})$ and $%
Z_{\alpha }Z_{j}(-\theta _{2})Z_{i}(\theta _{1})$ leads to 
\begin{eqnarray}
R_{i\alpha }^{k\beta }(\theta _{1})\tilde{R}_{j\beta }^{l\gamma }(\theta
_{2}) &=&R_{i\beta }^{k\gamma }(\theta _{1})\tilde{R}_{j\alpha }^{l\beta
}(\theta _{2}),  \label{YBII1} \\
\tilde{R}_{l\beta }^{q\gamma }(\theta _{2})S_{kj}^{lp}(\hat{\theta}%
_{12})T_{i\alpha }^{k\beta }(\theta _{1}) &=&S_{lk}^{qp}(\theta
_{12})T_{i\beta }^{l\gamma }(\theta _{1})\tilde{R}_{j\alpha }^{k\beta
}(\theta _{2}),  \label{YBII2} \\
S_{lq}^{st}(\theta _{12})R_{p\beta }^{q\gamma }(\theta _{1})S_{ik}^{lp}(\hat{%
\theta}_{12})\tilde{T}_{j\alpha }^{k\beta }(\theta _{2}) &=&R_{i\alpha
}^{s\beta }(\theta _{1})\tilde{T}_{j\beta }^{t\gamma }(\theta _{2}),
\label{YBII3} \\
\tilde{T}_{l\beta }^{q\gamma }(\theta _{2})S_{kj}^{lp}(\hat{\theta}%
_{12})T_{i\alpha }^{k\beta }(\theta _{1}) &=&T_{t\beta }^{p\gamma }(\theta
_{1})S_{ik}^{qt}(\hat{\theta}_{12})\tilde{T}_{j\alpha }^{k\beta }(\theta
_{2}).  \label{YBII4}
\end{eqnarray}

Clearly the three sets of Yang-Baxter equations (\ref{YB1})-(\ref{YB4}), (%
\ref{YB1})-(\ref{YB4}) with $R\leftrightarrow \tilde{R}$ and $%
T\leftrightarrow \tilde{T}$ and (\ref{YBII1})-(\ref{YBII4}) are not
equivalent. In the special case when $R=0$ the equation (\ref{YB4}) can be
turned into (\ref{YBII4}) by means of the unitarity relation (\ref{U2}). On
the other hand, when $T=0$ the equations (\ref{YBII1}) remain a non-trivial
requirement which links the left and right reflection amplitude via the
impurity degrees of freedom.

In order to achieve a more concise formulation, let us re-write the defect
Yang-Baxter equations in tensor form in the bulk indices. Employing the
usual convention $(A\otimes B)_{ij}^{kl}=A_{i}^{k}B_{j}^{l}$ for the tensor
product, the three non-equivalent equations in (\ref{YB1})-(\ref{YB4}) take
on the form 
\begin{eqnarray}
S(\theta _{12})[\Bbb{I}\otimes R_{\alpha }^{\beta }(\theta _{1})]S(\hat{%
\theta}_{12})[\Bbb{I}\otimes R_{\beta }^{\gamma }(\theta _{2})]\!\!\!\!
&=&\!\!\!\![\Bbb{I}\otimes R_{\alpha }^{\beta }(\theta _{2})]S(\hat{\theta}%
_{12})[\Bbb{I}\otimes R_{\beta }^{\gamma }(\theta _{1})]S(\theta
_{12}),\,\,\,\,\,\,\,  \label{YBt1} \\
S(\theta _{12})[\Bbb{I}\otimes R_{\alpha }^{\beta }(\theta _{1})]S(\hat{%
\theta}_{12})[\Bbb{I}\otimes T_{\beta }^{\gamma }(\theta _{2})]\!\!\!\!
&=&\!\!\!\!R_{\beta }^{\gamma }(\theta _{1})\otimes T_{\alpha }^{\beta
}(\theta _{2}),  \label{YBt2} \\
S(\theta _{12})[T_{\alpha }^{\beta }(\theta _{2})\otimes T_{\beta }^{\gamma
}(\theta _{1})]\!\!\!\! &=&\!\!\!\![T_{\alpha }^{\beta }(\theta _{1})\otimes
T_{\beta }^{\gamma }(\theta _{2})]S(\theta _{12}),  \label{YBt3}
\end{eqnarray}
whereas (\ref{YBII1})-(\ref{YBII4}) can be equivalently written as 
\begin{eqnarray}
R_{\alpha }^{\beta }(\theta _{1})\otimes \tilde{R}_{\beta }^{\gamma }(\theta
_{2}) &=&R_{\beta }^{\gamma }(\theta _{1})\otimes \tilde{R}_{\alpha }^{\beta
}(\theta _{2}),  \label{RR} \\
\lbrack T_{\alpha }^{\beta }(\theta _{2})\otimes \Bbb{I}]S(\hat{\theta}%
_{12})[\tilde{R}_{\beta }^{\gamma }(\theta _{1})\otimes \Bbb{I}]S(\theta
_{12}) &=&T_{\beta }^{\gamma }(\theta _{2})\otimes \tilde{R}_{\alpha
}^{\beta }(\theta _{1}),  \label{TR} \\
\lbrack \Bbb{I}\otimes \tilde{T}_{\alpha }^{\beta }(\theta _{2})]S(\hat{%
\theta}_{12})[\Bbb{I}\otimes R_{\beta }^{\gamma }(\theta _{1})]S(\theta
_{12}) &=&R_{\alpha }^{\beta }(\theta _{1})\otimes \tilde{T}_{\beta
}^{\gamma }(\theta _{2}),  \label{RT} \\
\lbrack T_{\alpha }^{\beta }(\theta _{1})\otimes \Bbb{I}]S(\hat{\theta}%
_{12})[\tilde{T}_{\beta }^{\gamma }(\theta _{2})\otimes \Bbb{I}] &=&[\Bbb{I}%
\otimes \tilde{T}_{\alpha }^{\beta }(\theta _{2})]S(\hat{\theta}_{12})[\Bbb{I%
}\otimes T_{\beta }^{\gamma }(\theta _{1})].
\end{eqnarray}
Making now the following assumption\footnote{%
Similar conclusions can be drawn by presuming 
\[
\tilde{T}_{\alpha }^{\beta }(\theta _{1})\otimes \tilde{R}_{\beta }^{\gamma
}(\theta _{2})=\tilde{T}_{\beta }^{\gamma }(\theta _{1})\otimes \tilde{R}%
_{\alpha }^{\beta }(\theta _{2})
\]
from the Yang-Baxter equations (\ref{YBt2}) with $T\rightarrow \tilde{T}$, $%
R\rightarrow \tilde{R}$. Alternatively, assuming 
\[
T_{\alpha }^{\beta }(\theta _{1})\otimes \tilde{R}_{\beta }^{\gamma }(\theta
_{2})=T_{\beta }^{\gamma }(\theta _{1})\otimes \tilde{R}_{\alpha }^{\beta
}(\theta _{2})
\]
or 
\[
\tilde{T}_{\alpha }^{\beta }(\theta _{1})\otimes R_{\beta }^{\gamma }(\theta
_{2})=\tilde{T}_{\beta }^{\gamma }(\theta _{1})\otimes R_{\alpha }^{\beta
}(\theta _{2})
\]
we can draw the same conclusions from (\ref{TR}) or (\ref{RT}),
respectively. Notice that only one of these four assumptions is sufficient.
Furthermore, it is enough if only one of the matrices involved in the
product is diagonal (abelian) in the impurity degrees of freedom. This
implies of course that the other matrix can be completely generic.} on the
product of $R$ and $T$ in the impurity degrees of freedom 
\begin{equation}
T_{\alpha }^{\beta }(\theta _{1})\otimes R_{\beta }^{\gamma }(\theta
_{2})=T_{\beta }^{\gamma }(\theta _{1})\otimes R_{\alpha }^{\beta }(\theta
_{2})\,.  \label{ass}
\end{equation}
we can for instance eliminate $T(\theta )$ in (\ref{YBt2}). Taking
thereafter $\theta _{2}=0$, we obtain 
\begin{equation}
S(\theta )\left[ \Bbb{I}\otimes R_{\alpha }^{\beta }(\theta )\right]
S(\theta )=R_{\alpha }^{\beta }(\theta )\otimes \Bbb{I\,}\,.  \label{eq}
\end{equation}
Therefore it follows by relativistic invariance that the scattering matrix
has to be rapidity independent, i.e. a constant matrix that is of the form 
\begin{equation}
S(\theta )\Bbb{=P}\sigma \,,  \label{S}
\end{equation}
where $\Bbb{P}$ is a permutation operator and $\sigma $ a constant matrix.
One may of course reverse the argument and take $\sigma $ to be a phase with
properties $\sigma _{ij}\sigma _{ji}=1$, substitute (\ref{S}) back into (\ref
{YBt2}) and deduce (\ref{ass}). Obviously similar conclusions can be reached
when solving (\ref{YBt2}) for $S(\hat{\theta}_{12})$ or when considering (%
\ref{TR}), (\ref{RT}) in the same manner.

In summary: \emph{Apart from rapidity independent scattering matrices,
simultaneous reflection and transmission in an integrable system with an
impurity is always absent when there are no degrees of freedom in the
defect. If there are degrees of freedom in the defect, reflection and
transmission can only occur for non-diagonal bulk theories when (\ref{ass})
does not hold. }

\section{Defect bootstrap equations}

Besides exploiting the associativity of the ZF-algebra in the above version,
the presence of bound states in the bulk theory also leads to powerful
constraints. Despite the fact that equation (\ref{S}) already manifests that 
$S$ is independent of the rapidity, let us
exploit the associativity of the expression which reflects this situation 
\begin{equation}
Z_{a}\left( \theta +i\eta _{ac}^{b}+i\varepsilon /2\right) Z_{b}\left(
\theta -i\eta _{bc}^{a}-i\varepsilon /2\right) =i\Gamma _{ab}^{c}Z_{c}\left(
\theta \right) /\varepsilon \,,
\end{equation}
for $\varepsilon \rightarrow 0$. As conventional we denote here the three
particle vertex on mass-shell by $\Gamma _{ab}^{c}$ and the real fusing
angles by $\eta $.

Commuting then in the manner as depicted in figure 2 leads to non-trivial
constraints for the defect scattering matrices. This means scattering the
particles $a$ and $b$ on the defect and fusing afterwards to particle $c$
should be equivalent to fusing first to particle $c$ and scatter thereafter
onto the defect. We end up with the following sets of equations 
\begin{eqnarray}
R_{a}\left( \theta +i\eta _{ac}^{b}\right) R_{b}\left( \theta -i\eta
_{bc}^{a}\right) S_{ab}(2\theta +i\eta _{ac}^{b}-i\eta _{bc}^{a})
&=&R_{c}\left( \theta \right) \,,  \label{fu1} \\
T_{a}\left( \theta +i\eta _{ac}^{b}\right) T_{b}\left( \theta -i\eta
_{bc}^{a}\right) &=&T_{c}\left( \theta \right) \,,  \label{fu2} \\
T_{a}\left( \theta +i\eta _{ac}^{b}\right) R_{b}\left( \theta -i\eta
_{bc}^{a}\right) S_{ab}(\theta +i\eta _{ac}^{b}-i\eta _{bc}^{a}) &=&0\,,
\label{fu3} \\
R_{a}\left( \theta +i\eta _{ac}^{b}\right) T_{b}\left( \theta -i\eta
_{bc}^{a}\right) &=&0\,,  \label{fu4}
\end{eqnarray}
where we suppressed the explicit mentioning of the degrees of freedom of the
impurity. Obviously we can derive the same equations also for $%
R\leftrightarrow \tilde{R}$ and $T\leftrightarrow \tilde{T}$. It is evident
that the equations (\ref{fu1})-(\ref{fu4}) only make sense when either $T=0$
or $R=0$, in which case the reflection bootstrap equation (\ref{fu1})
(proposed first in \cite{FK}) and the transmission bootstrap (\ref{fu2})
(proposed first in \cite{Konik}) separately become meaningful. Hence, we
have confirmed in an alternative way a statement which already followed from
the previous section, namely: \textit{A bulk theory which possesses bound
states associated with a diagonal scattering matrix does not allow
simultaneously non-vanishing transmission and reflection through a defect. }%
The argument leading to the equations (\ref{fu1})-(\ref{fu4}) gives a
slightly more intuitive understanding for the exclusion of this possibility,
since it shows that one produces inevitably terms made up of particles on
the left and on the right of the defect which can not be reconciled anymore
such that (\ref{fu3}) and (\ref{fu4}) have to hold. Nonetheless, one should
note that equation (\ref{S}) is more restrictive since it also excludes
theories which do not permit fusing at all, such as the sinh-Gordon model,
etc.

\begin{center}
\includegraphics[width=11.2cm,height=7.77cm]{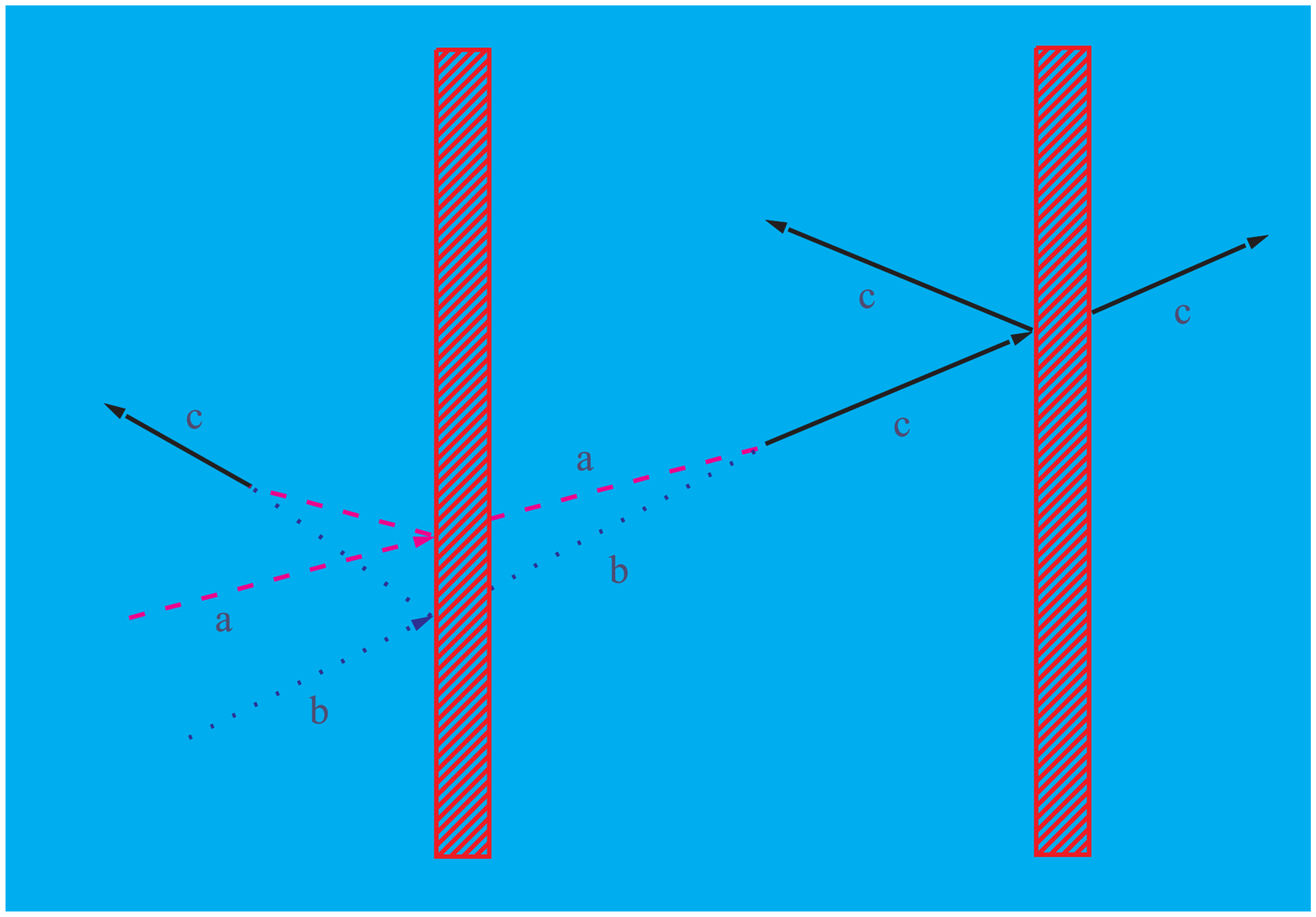}
\end{center}

\noindent {\small Figure 2: Defect bootstrap equations.} \medskip

Let us now consider the possibility of impurity excitations and the related
bootstrap equations. Having a particle of type $a$ moving at rapidity $i\eta
_{a\alpha }^{\beta }$ might change the state of the defect from $\alpha $ to 
$\beta $ as 
\begin{equation}
Z_{a}\left( i\eta _{a\alpha }^{\beta }\right) Z_{\alpha }\rightarrow
Z_{\beta }\,.
\end{equation}
Using this relation and commuting in accordance with the ZF-algebra (\ref{Z1}%
)-(\ref{Z2}) in an order as depicted in figure 3 leads to a set of defect
bootstrap equations involving the degrees of freedom of the impurity 
\begin{eqnarray}
R_{b\beta }(\theta ) &=&S_{ab}\left( \theta -i\eta _{a\alpha }^{\beta
}\right) S_{ba}\left( \theta +i\eta _{a\alpha }^{\beta }\right) R_{b\alpha
}(\theta )\,,  \label{bb1} \\
T_{b\beta }(\theta ) &=&T_{b\alpha }(\theta )\,.  \label{bb2}
\end{eqnarray}
It follows trivially from (\ref{bb2}) that whenever the transmission is
non-vanishing there can not be any excited impurity states. In that case
this is compatible with (\ref{bb1}), since $S$ is a constant phase which
cancels due to the unitarity relation (\ref{U1}). On the other hand,
whenever $T(\theta )$ is zero, equation (\ref{bb1}) should be satisfied and
can be used for the construction of new solutions for $R(\theta )$. The
possible first order poles in $R(\theta )$, related to the fusing angles,
have to be in the physical sheet, i.e., $0<\func{Im}\theta <\pi $, and
should be associated with a positive residue \cite{FK2}. Using these
criteria one may find non-trivial closures of the boundary bound state
bootstrap equation \cite{FK2,Riva}. Reversing this statement means, of
course, that in a consistent solution for $R(\theta )$ and $T(\theta )$
every pole inside the physical sheet should be related to a negative
residue. Let us verify this with explicit examples.

\begin{center}
\includegraphics[width=11.2cm,height=7.77cm]{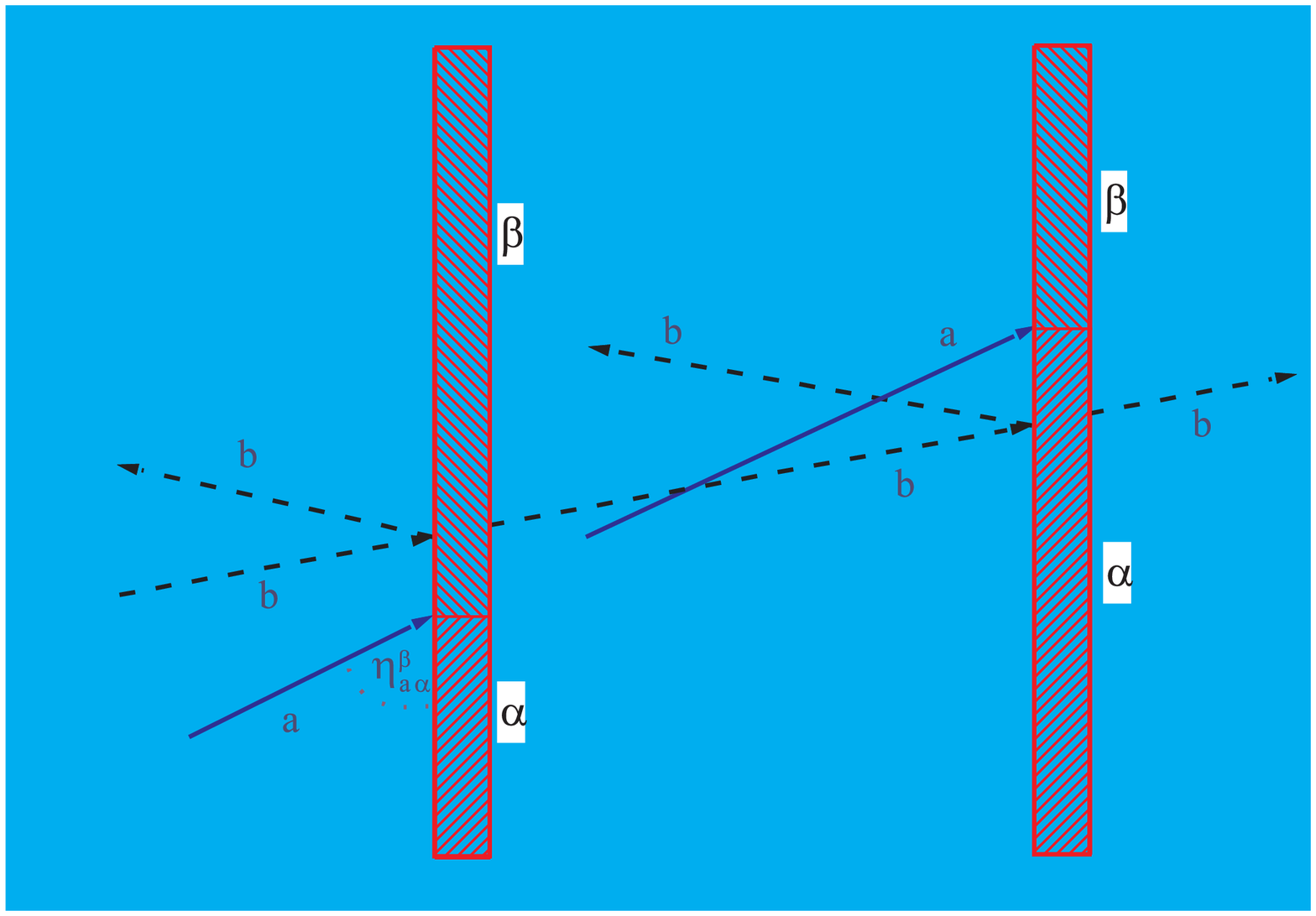}
\end{center}

\noindent {\small Figure 3: Defect bound state bootstrap equations.} \medskip

\section{The free Fermion with a defect}

We consider the complex free Fermion Lagrangian density $\mathcal{L}_{FF}$
perturbed with a defect $\mathcal{D}(\bar{\psi},\psi )$ 
\begin{equation}
\mathcal{L}=\,\mathcal{L}_{FF}+\delta (x)\mathcal{D}(\bar{\psi},\psi )\,.
\label{La}
\end{equation}
Here we denote as usual $\bar{\psi}=\psi ^{\dagger }\gamma ^{0}$, where $%
\gamma ^{0}$ is one of the gamma matrices, i.e., satisfying the Clifford
algebra. For the defect $\mathcal{D}(\bar{\psi},\psi )=g\bar{\psi}\psi $ the
transmission and reflection amplitudes were computed \cite{DMS,CF10} to be 
\begin{eqnarray}
R(\theta ,B) &=&\bar{R}(\theta ,-B)=-\frac{i\sin B\cosh \theta }{\sinh
\theta +i\sin B}\,,  \label{t1} \\
T(\theta ,B) &=&\bar{T}(\theta ,-B)=\frac{\cos B\sinh \theta }{\sinh \theta
+i\sin B}\,.  \label{t2}
\end{eqnarray}
Since Dirac Fermions are not self-conjugate, we have to distinguish particle
and anti-particle. We denote the amplitudes related to the anti-particle by
a ``bar''. The coupling constant $g$ is parameterized as $\sin
B=-4g/(4+g^{2})$. For this example parity invariance is preserved, such that 
$R=\tilde{R}$ and $T=\tilde{T}$.\ The fact that 
\begin{equation}
\limfunc{Res}\limits_{\theta \rightarrow -iB}R(\theta ,B)=\limfunc{Res}%
\limits_{\theta \rightarrow -iB}T(\theta ,B)=-\limfunc{Res}\limits_{\theta
\rightarrow iB}\bar{R}(\theta ,B)=-\limfunc{Res}\limits_{\theta \rightarrow
iB}\bar{T}(\theta ,B)=2\pi \sin B\,,  \label{res}
\end{equation}
confirms our previous conclusion, which asserted that there can not be any
excited impurity states once reflection and transmission occur
simultaneously. Depending on the sign of $B$, the residues in (\ref{res})
are either negative or the pole is beyond the physical sheet. Thus, this
solution is consistent with regard to the above argumentation. This is in
contradiction to the statements in \cite{DMS}, where it was argued that
excited impurity states do exist.

As the second example we consider the defect $\mathcal{D}(\bar{\psi},\psi )=g%
\bar{\psi}\gamma ^{0}\psi $ for which the related transmission and
reflection amplitudes follow from \cite{CF10} as 
\begin{eqnarray}
R(\theta ,B) &=&\bar{R}(\theta ,B)=\frac{-i\sin B}{\sinh (\theta +iB)}\,, \\
T(\theta ,B) &=&\bar{T}(\theta ,B)=\,\frac{\sinh \theta }{\sinh (\theta +iB)}%
\,.
\end{eqnarray}
Also in \ this example parity invariance is preserved, and we have $R=\tilde{%
R}$ and $T=\tilde{T}$.\ We compute 
\begin{equation}
\limfunc{Res}\limits_{\theta \rightarrow -iB}R(\theta ,B)=\limfunc{Res}%
\limits_{\theta \rightarrow -iB}T(\theta ,B)=2\pi \sin B\,,
\end{equation}
such that the interpretation is the same as in the previous example and we
confirm once more our general statement.

\section{Conclusions}

We conclude by re-stating our main results: \emph{When there are no degrees
of freedom in the defect, the only integrable, in the sense of
factorization, bulk theories which, when doped with some impurity, which
allow the occurrence of simultaneous reflection and transmission are those
possessing constant, i.e. rapidity independent, scattering matrices. Once }$T
$ \emph{and }$R$ \emph{are taken simultaneously to be non-vanishing these
theories do not admit the possibility of excited impurity bound states. For
this to happen in complete generality (\ref{ass}) has to be violated.   }

Unfortunately so far our main results imply that for the treatment of
non-trivial impurity systems one has to leave the realm of integrable
systems. It would be interesting to construct $T$ and $R$ for some
non-diagonal bulk theories, by means of (\ref{YB1})-(\ref{YB4}) and (\ref
{YBII1})-(\ref{YBII4}) under the violation of (\ref{ass}). For the subset of
diagonal bulk theories which remain integrable in this case, the restrictive
power of the integrability framework fails, since apart from crossing and
unitarity we have no constraining equations at our disposal to determine the
transmission and reflection amplitudes. It would be interesting to complete
the picture for these anyonic theories and seek also solutions for the
Federbush type models \cite{Feder,Fform}.\medskip 

\noindent \textbf{Acknowledgments: } O.A. C-A. and A.F. are grateful to the
Deutsche Forschungsgemeinschaft (Sfb288) for financial support.

\end{document}